# Electrothermally modulated stop-go microvalves for capillary transport systems

Golak Kunti,[1] Debabrata DasGupta,[2] Anandaroop Bhattacharya,[1] & Suman Chakraborty[1, a]

[1]Department of Mechanical Engineering, Indian Institute of Technology Kharagpur, Kharagpur, West Bengal - 721302, India

[2]Department of Mechanical Engineering, Indian Institute of Technology Delhi, New Delhi-110016, India

[a]*E-mail address* of corresponding author: suman@mech.iitkgp.ernet.in

**ABSTRACT**

In this article, we show a novel approach of implementing valving action in a dynamically evolving capillary filling process, by exploiting alternating current electrothermal (ACET) mechanism. The surfaces of the top and bottom walls of the capillary are asymmetrically patched with two different wettabilities. Our observations reveal that by exploiting an intricate interplay of the electrothermal effects and interfacial tension modulation, the acceleration-deceleration motion of the interface and interface pinning at specific locations may be delicately tuned. This, in turn, modulates the flow speed and residence time of the interface at designated locations, with a simple tuning of the frequency of the electrical signal. The present study, thus, unveils a novel microvalving mechanism achieved via effective maneuvering of the interfacial electrochemistry where active and passive control mechanisms can be utilized simultaneously to achieve control over the interfacial motion of the two-phase system. This may bear far reaching consequence in optimizing several applications in the bio-microfluidic domain, including devising novel miniaturized platforms for 'locally' investigating reaction chemistries.

**I. INTRODUCTION**

Investigations on interfacial motion of immiscible fluids are important in various natural processes, and have widespread applications in the fields of engineering and applied sciences [1–3]. Bio-chemical analysis, biomedical processes, chemical technologies, oil recovery, textile manufacturing, polymer processing, manufacturing of photographic films are some of the important areas where such interfacial flows may play critical roles [4–9]. Accordingly, various studies have been directed towards understanding the underlying fundamental mechanisms governing the transport characteristics of multiphase microfluidics from the perspective of contact line dynamics [10–12], or specifically focused towards addressing particular applications such as bio-chemical reactions [13], particle deposition [14], clinical diagnostics [15], biochemical analysis [16], drug delivery [17], chemical synthesis [18], heat transfer coating [19] etc.

Capillarity arises when fluids wet the surface of the channel and effect of the surface tension force becomes dominating. The free energy of the surfaces is deeply influenced by the surface patterning. Alternating periodic patches have strong effects on tuning of flow behavior and interfacial dynamics of binary fluids [20,21]. Periodic alteration of the surface free-energy, when the interface traverses from one surface patch to another (e.g., hydrophobic and hydrophilic stripes), changes surface tension force and thereby alters the shape of the interface of the immiscible fluid system [21,22]. Therefore, surface patterning can be engineered/tuned depending on the applications, for example, controlling of flow field [23,24], imposing flow instability [25] etc. Moreover, wetting and dewetting of the binary fluids over patterned surfaces may lead to interface breaking event at high diffusion rates owing to high wettability difference between wettability patches [21]. Several other aspects of utilization of surface patterning to control the fluid flow can be found in the Refs. [2,26,27] Texturing over substrates can be made using the following techniques: micro-photolithography [28], self-assembling monolayers (SAM) [29], adsorption of polymer [30], alteration of physico-chemical properties of surface [31,32], multistream liquid laminar flow [2], plasma treatment [33], to name a few. Recently, directional wetting i.e., capillary wicking has emerged as an alternative to the traditional means of liquid motion control [34,35].

Handling and manipulation of small volumes of liquid inside a microchannel are the key requisites for various biological and chemical processes [36–39]. Various techniques, such as transporting, metering and positioning to control the fluid motion are based on mechanical pumping, electrowetting, magnetic and electric field driven pumping etc [40–44]. The aforementioned methodologies are active control-based methods where a very small amount of liquid can be precisely and rapidly handled. In recent years, electrical actuators are playing important role to actuate the flow and enhance the efficiency, portability and reliability of microfluidic devices. On-chip integrability, low noise, no complex moving part etc. are the crucial aspects which make the electrical actuators more sought after for miniaturized devices. Compared to DC electrical sources, AC sources possess several advantages, one key advantage being operation at low voltage ($<10$ V [45–47]). Therefore, the detrimental effects of DC sources viz. at high voltage Joule heating, formation of bubbles, electrolysis etc. can be avoided using AC voltages [48–50]. Under AC electric field, AC electroosmosis (ACEO) force arises because of mobile free charges in the double layer over the electrode surfaces [51,52]. Besides, AC electrothermal (ACET) forces are caused by induced charge owing to nonuniform AC electric field and inhomogeneities in permittivity and electrical conductivity caused by temperature gradient [52–56]. ACEO process operates at low frequency (up to 100 KHz) and electrical conductivity ($\leq 1$ mS/m) [57]. However, ACEO forces are drastically reduced at relatively high frequency (>100 kHz [58,59]) and solution conductivity ($>0.085$ S/m [60]). On the other hand, ACET mechanism operates over a wide range of operating frequencies and electrical conductivities thereby making them an attractive alternative [39,45,55,61].

In modern days, reduced volume, portability, reliability and shortened investigation time have promoted the usage of microfluidic devices for chemical and biomedical analysis [62–64]. However, several critical aspects limit the applicability of the devices in those applications. One of the major concern is flow control inside a microchannel where microvalves are required. Conventional microvalves deal with mechanical [65], magnetic [66], and pneumatic [67] forces. Microfluidic devices which use conventional valves involve moving and sealing parts, rendering these devices bulky. Nevertheless, some nonmechanical valves, such as ice microvalves [68], paraffin microvalves [69], hydrogel-based micro-valves [70] etc. can avoid the drawbacks of the mechanical valves. However, these non-mechanical valving actions are often premised on the alteration of important properties like pH, ionic strength, solvent composition etc., which, in turn, makes the integrated system much complicated for design and fabrication.

Here, we introduce a novel active (ACET) and/or passive (tuned wettability patterns) means of controlling the dynamics of two immiscible fluids in a capillary. Recently, ACET mechanism has been shown to be an efficient and effective means for fluid pumping, specifically for biological and biochemical applications [38,60,71]. On the other hand, by controlling the frequency of the AC signal, flow can be easily and rather effectively maneuvered [46,57,72]. In the present study, we bring out a non-trivial interplay between ACET forces and interfacial tension on a patterned microfluidic substrate towards dynamic tuning of capillary front progression in a confined environment. The surfaces of the capillary are patterned with two different wettability patches. The patterned patches on the top and bottom walls are different at same axial location. Therefore, while the patches paved on top wall attract one fluid (say fluid A), the patches paved on the bottom wall repel that fluid (i.e., fluid A) at same axial location. Also, it is possible to control the flow at any instant of time at any location by merely modulating the actuating electrical signal. We demonstrate that by appropriately tuning the various parameters of surface texturing and the relevant electrical parameters, it is possible to control the fluid flow and its residence time at specific locations along the channel, thereby realizing valving actions that are unique to capillaty driven miniaturized devices of these kinds.

## II. PROBLEM FORMULATION AND THEORY

### A. Physical system

The primary objective of the present study is to investigate the transport of a controlled binary fluid system through a narrow fluidic channel where AC electrothermal forces are used as a driving potential. The binary fluid system of fluids A and B is shown in the Fig. 1(a). We consider a channel of height and length of $H$ and $L$, respectively. The origin of the coordinate system is placed at left bottom corner of the domain, while $x$ axis and $y$ axis run along the length and height of the channel, respectively. The length along the third direction is large enough compared to the height and thus two-dimensional analysis is considered. The electrode structures are shown in the Fig. 1(b), We consider 10 pairs of asymmetric electrodes embedded over the top and bottom wall to get a net flow along the longitudinal direction. The various dimensions of the electrode structure are shown in the figure. The surfaces of the channel are paved with wettability patches which are manifested by imposed static contact angle $\theta_s$. Two different kinds of wetting condition $\theta_A$ and $\theta_B$ are imposed on the patches periodically from a distance of $L_1$ and spans over a distance of $L_2$. We consider asymmetric patterning of the patches with respect to top and bottom walls where axial locations of the patches are interchanged. Over the top wall we consider the characteristics of the patches as:

$$\theta_s = \{\theta_A \forall (2n-1)p_w \le x' \le 2np_w,$$
$$= \{\theta_B \forall 2(n-1)p_w \le x' \le (2n-1)p_w, \tag{1}$$

Whereas for bottom wall the characteristics of the patches are

$$\theta_s = \{\theta_A \forall 2(n-1)p_w \le x' \le (2n-1)p_w,$$
$$= \{\theta_B \forall (2n-1)p_w \le x' \le 2np_w, \tag{2}$$

where $x' = x - L_1$ and $n$ is the number of patches of one kind. $p_w$ is the patch width. Further, we consider $\theta_s = \pi/2$ for remaining portions of the walls. Initially, immiscible fluids A and B reside at the left and right portions of the channel, respectively. Fluids start moving when electrothermal forces are generated in the solution domain on application of voltage on the electrodes. The details of the numerical modeling are discussed in the subsequent section.

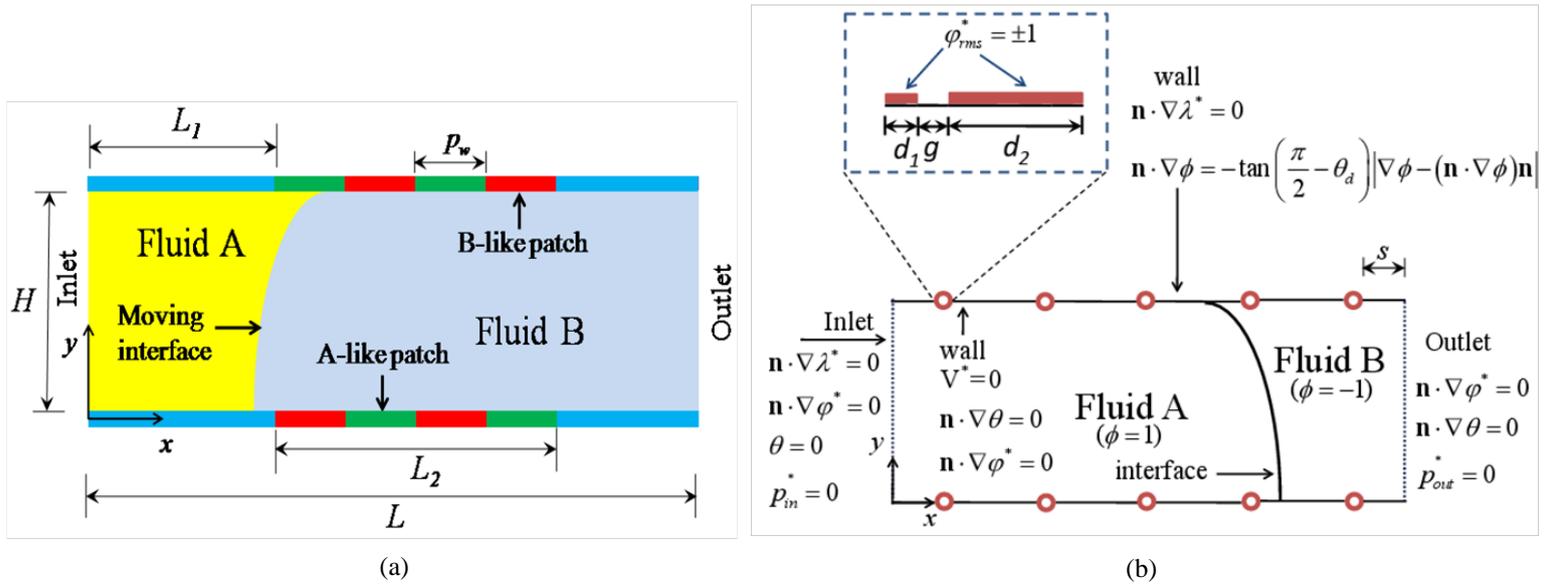

FIG. 1. (a) Physical system of the present study. The origin is placed at the left bottom corner of the domain. Fluid A (yellow color) and fluid B (grey color) occupy the left and right portions of the solution domain respectively while walls are paved with two different wettability patches A-like patches (green color) and B-like patches (red color). Axial positions of the patches (A and B patches) are different for bottom and top walls which make the patches asymmetric with respect to the two walls (b) All boundary conditions adopted in the simulations. Electric field is applied by imposing AC voltages on the electrodes. Magnified view of one pair of electrode is shown in the dotted box. On application of electrical signal ACET forces will be generated and fluid starts moving along the longitudinal direction.

**B. Numerical modeling and mathematical formulation**

Phase field method is adopted to consider the interfacial coupling of the two-phase fluids. The equilibrium properties are determined by the free energy function which is described by order parameter $\phi$. The order parameter is uniform throughout the bulk phases while it gets altered at the diffused interface. The concentration of each phase is evaluated by the order parameter. We consider $\phi = 1$ and $\phi = -1$ for fluid A and B, respectively, while at the interface it changes its value from +1 to -1, $\phi = 0$ being the nominal location of the interface separating the two fluids. The thermodynamics of the binary system is governed by the Ginzburg-Landau free energy functional [32,73–77]:

$$F = \int_\forall \left\{ f(\phi) + \frac{1}{2}\eta\xi|\nabla\phi|^2 \right\} d\forall, \tag{3}$$

where, $\forall$ is the volume of the domain, interface thickness is $\xi\,(=0.025H)$ and $\eta$ is the surface tension. The first term in the integral, $f(\phi)$, is the bulk free energy density. The second term in the integral takes into account of the excess free energy owing to the presence of the diffuse interface separating the phases. The total bulk free energy $f(\phi)$ can be written in terms of double-well potential as $f(\phi) = \frac{\eta}{4\xi}(1-\phi^2)^2$ [78–80]. The minimum value of the total bulk free energy can be obtained for $\phi = \pm 1$. The chemical potential can be expressed as $\lambda = \frac{\delta F}{\delta \phi} = f'(\phi) - \eta\xi\nabla^2\phi$. The nondimensional form of the chemical potential and Cahn-Hilliard equation (CHE) which governs the evolution of order parameter $\phi$ is expressed by [78,79]

$$\lambda^* = -\mathrm{Cn}\,\nabla^2\phi + \frac{1}{\mathrm{Cn}}(\phi^3 - \phi), \tag{4}$$

$$\frac{D\phi}{Dt^*} = \frac{1}{\mathrm{Pe}_\phi}\nabla^2\lambda^*, \tag{5}$$

where $Cn = \xi/H$ is the Cahn number and $Pe_\phi = u_0 H^2 / M_c \eta$ is the phase field Peclet number. The dimensionless mobility parameter is $M^* = M/M_c$, where $M\,(>0)$ is the mobility parameter. The mobility parameter controls the interfacial diffusion rate. $M_c\,(= Cl^4/\sqrt{\tilde{m}\varepsilon_e})$ is defined following the MD simulation studies of Qian et al. [81]. $C\,(= 0.023)$ is a constant, $\tilde{m}$ is the molecular mass, $l$ and $\varepsilon_e$ are the length scale and energy scale of the Linnard-Jones potential, respectively. The channel height ($H$) is taken as reference length and properties of the fluid A are chosen as reference fluid properties. The nondimensional parameters are represented with superscript *. Other nondimensional parameters are $t^* = \frac{tu_0}{H}, \theta = \frac{T-T_0}{\Delta T}, \varphi^* = \frac{\varphi}{\varphi_0}, p^* = \frac{pH}{\mu u_0}, \lambda^* = \frac{\lambda}{\eta/H}$, and $\omega^* = \frac{\omega}{\omega_0}$. Here, $t, T, \varphi, p,$ and $\omega$ are the time, temperature, potential, pressure, and applied frequency, respectively. The reference velocity is $u_0 = \varepsilon_0 \varphi_0^2 \Delta T \beta_0 / \mu_0 H$, $\Delta T$ being the maximum temperature difference. In an electrothermal domain, the maximum temperature occurs at the gap between the electrodes. In our analysis, we consider initial temperature of the fluid and reference temperature to be the same. To estimate the order of reference velocity we use the following values which are best fitted for aqueous solution: Permittivity $\varepsilon_0 \approx 10^{-10}\,\text{C/Vm}$, $\varphi_0 \approx 1\,\text{V}$, $\Delta T \approx 1\,\text{K}$, $\beta_0 \approx 0.01\,\text{K}^{-1}$, $\mu_0 \approx 0.001\,\text{Pa s}$ and $H \approx 10^{-4}\,\text{m}$. The order of reference velocity is $u_0 \approx 10^{-5}\,\text{m/s}$. Aqueous KCl solution is adopted as base fluid for which gradient of electrical conductivity $\beta_0$ and gradient of the permittivity ($\alpha_0$) are of the order of 0.01 and -0.001, respectively [82].

We consider the electric field to be quasi-electrostatic such that the induced magnetic field can be neglected [83,84]. In an AC field, electrical potential can be obtained solving the following equation [85]

$$\nabla \cdot \left( \sigma^* \nabla \varphi^* \right) = 0 \tag{6}$$

where $\sigma^*$ is the electrical conductivity. The externally applied AC electric field causes heat generation in the bulk fluid domain. This heat source, also known as Joule heat, causes temperature distribution as evident from the energy equation

$$\text{Pe}_T \, \rho^* C_p^* \frac{D\theta}{Dt^*} = k^* \nabla^2 \theta + \frac{1}{2} J \sigma^* \left| \mathbf{E}^* \right|^2, \tag{7}$$

where $C_p^*$ is the specific heat and $k^*$ is the thermal conductivity. $\sigma^* \left| \mathbf{E}^* \right|^2$ is the Joule heat. $Pe_T = \rho_0 C_{p0} u_0 H / k_0$ is the thermal Peclet number and $J = \sigma_0 \phi^2 / k_0 \Delta T$ is the Joule number. $\rho_0$ is th density of the fluid A.

The velocity field is obtained by solving the Navier-Stokes equation. Present analysis deals with an electric field driven actuation mechanism where thermal field has strong effect to cause the electrothermal forces. In addition to the electrothermal forces, surface force and thermocapillary force may interplay into the two phase system. Considering laminar and incompressible fluid flow, the modified continuity equation and Cahn-Hilliard-Navier-Stokes equation in nondimensional form can be written as [21,86,87]

$$\nabla \cdot \mathbf{V}^* = 0, \tag{8}$$

$$\text{Re}\,\rho^* \frac{D\mathbf{V}^*}{Dt^*} = -\nabla p^* + \nabla \cdot \left[ \mu^* \left( \nabla \mathbf{V}^* + \nabla \mathbf{V}^{*T} \right) \right] + \frac{1}{\text{Ca}} \mathbf{F}_\mathbf{S}^* + \zeta \mathbf{F}_\mathbf{E}^*, \tag{9}$$

Here, $\mathbf{V}^*$ is the fluid velocity and $\mu^*$ is the viscosity of the fluid. The dimensionless number $\mathrm{Re} = \rho_0 u_0 H / \mu_0$, $\mathrm{Ca} = \mu_0 u_0 / \eta$, and $\zeta == \varepsilon_0 \phi^2 / \mu_0 u_0 H$ are the Reynolds number, capillary number:, and ACET force number, respectively. The expression of the ACET force number is [52,88]

$$\mathbf{F_E} = -\frac{1}{2}\left[\left(\frac{\nabla\sigma}{\sigma} - \frac{\nabla\varepsilon}{\varepsilon}\right)\cdot\mathbf{E}\frac{\varepsilon\mathbf{E}}{1+(\omega\tau)^2} + \frac{1}{2}|\mathbf{E}|^2\nabla\varepsilon\right]. \tag{10}$$

where $\mathbf{E}(=-\nabla\varphi)$ is the electric field and $\tau = \varepsilon/\sigma$ is the charge relaxation time. The first term of the right hand side of the Eq. (10) is known as Coulomb force whereas second term is the dielectric force of the electrothermal forces. Combined effects of surface tension and thermocapillary forces are taken into account by the interfacial force $\mathbf{F_S} = \frac{3\sqrt{2}}{4}\xi\left[\eta_T|\nabla\phi|^2\nabla T - \eta_T(\nabla T \cdot \nabla\phi)\nabla\phi + \frac{\eta}{\xi^2}\lambda\nabla\phi\right]$ [89], where $\eta_T$ is the gradient of the interfacial tension. We consider a linear variation in surface tension with temperature which follows $\eta(T) = \eta_0 + \eta_T(T-T_0)$.

Various interfacial properties $\rho, k, C_p$ and $\mu$ are function of order parameter and can be casted in the form $f = 0.5 f_A(1+\phi) + 0.5 f_B(1-\phi)$ where $f$ may be $\rho, k, C_p$ or $\mu$. Electrothermal forces originate from the gradient of the electrical conductivity and permittivity which arise from the temperature gradient in the AC domain. Therefore, to consider temperature dependent conductivity and permittivity a linear variation of $\varepsilon$ and $\sigma$ with temperature is adopted. On the other hand, in phase-field formalism all the fluid properties are functions of the order parameter. Thus, expression of the electrical conductivity and permittivity with temperature and order parameter can be written by $f(T,\phi) = f_{0,A}\{1+b_A(T-T_0)\}\left(\frac{1+\phi}{2}\right) + f_{0,B}\{1+b_B(T-T_0)\}\left(\frac{1-\phi}{2}\right)$, where $f$ may be $\sigma$ or $\varepsilon$ and the gradient $b$ may be $\beta$ or $\alpha$.

To solve the CH equation, following two boundary conditions are adopted on the top and bottom walls:

$$\mathbf{n}\cdot\nabla\lambda = 0, \tag{11}$$

$$\mathbf{n}\cdot\nabla\phi = -\tan\left(\frac{\pi}{2}-\theta_d\right)|\nabla\phi - (\mathbf{n}\cdot\nabla\phi)\mathbf{n}|, \tag{12}$$

where $\mathbf{n}$ is the unit normal vector pointing outwards from the solid surface. $\theta_d$ is the dynamic contact angle. Eq. (11) implies zero flux through the solid surface whereas interface profile near the walls is modulated by the Eq. (12). [3] Important point to be mentioned that the consideration of the dynamic contact angle is taken through imposing static contact angle which is expressed by Cox–Voinov equation: [90,91] $\theta_d^3 = \theta_s^3 + 9Ca\ln(R/l_{slip})$, where R is the macroscopic length scale. $l_{slip}$ is the molecular slip length scale. To evaluate the capillary number in the above equation velocity in $Ca$ is taken as local contact line velocity and slip length is taken as interface thickness following the approach of Jacqmin [73]. Further, the applied voltages on the electrodes are $\varphi_{rms} = \pm 1$. The other boundaries are electrically insulated. To obtain the temperature distribution, the electrodes and inlet of the channel are set at reference temperature whereas at the outlet normal component of the heat flux is zero. The other boundaries are insulated. No penetration and no slip boundary conditions are imposed over the solid walls for the velocity field. Since, in the present analysis we consider electrically driven flow field, the gauge pressure at the inlet and outlet is set to zero. During initialization, the velocities and temperature are set as: $V^* = \theta = (x^*, y^*, t^* = 0) = 0 \,\forall\, x^*, y^*$.

## C. Numerical method, model benchmarking and grid independent study

AC electrothermally actuated flow of immiscible fluids over periodic pattern surfaces is a multi-phase, multi-physics phenomena. First, the electric field equation is solved to get the potential distribution over the domain. Next, remaining set of governing equations are solved. Finite Element Method based commercial software COMSOL is adopted to solve the dimensionless transport equations. For temporal discretization PARDISO solver and the generalized-$\alpha$ scheme are taken. For our analysis, free triangular meshes of sizes of $\Delta x^* = \Delta y^* = \xi$ are used. The details of the mesh independent study will be discussed in the next subsection. Initially a stable and equilibrium interface profile should be achieved before applying the electrical actuation to the fluid motion. This state will be obtained by initialization of the order parameter at the beginning of the numerical calculation and is performed by solving the equation: $\lambda(\phi) = 0$.

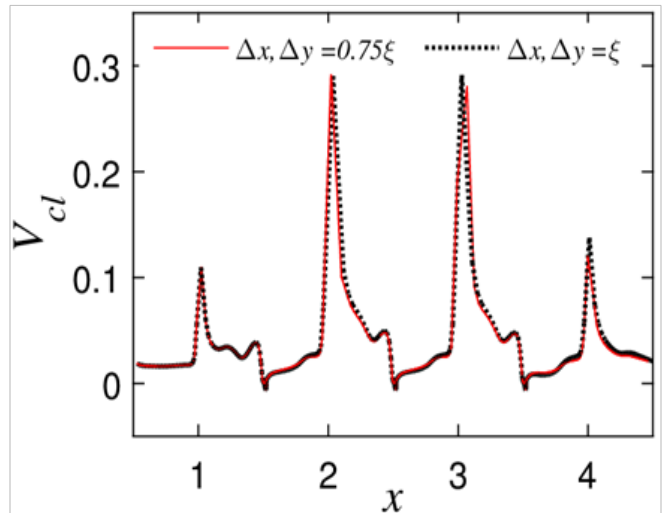

FIG. 2. shows mesh independent study of contact line velocity with contact line location for two different mesh sizes of $\Delta x, \Delta y = 0.75\xi$ and $\Delta x, \Delta y = \xi$. The parameters used in the results are $\theta_A = 60°$, $\theta_B = 120°$, $p_w = 0.5$, $\zeta = 50$ and $\omega = 0.001$. Negligible variation of the velocities is found for different mesh sizes.

We have benchmarked our numerical methodology using the results of Wang et al. [21] whose details can found in our previous work, Ref. [80] To achieve the sharp-interface limit and to avoid inaccuracy in the numerical results we have conducted mesh independent study. Fig. 2 depicts the results of the mesh independent study. We show the contact line velocity plot with axial distance over top wall, the relevant parameters used in the results are mentioned in the figure caption. When the interface crosses the junction between the two different stripes, sudden change in the free surface energy results in sudden change in contact line velocity. Hence, velocity variation shows abrupt increase from its mean position. The alteration of mesh sizes from $\Delta x, \Delta y = 0.75\xi$ to $\Delta x, \Delta y = \xi$, the result does not show appreciable difference in the variation of contact line velocities. So, in our investigation we use mesh sizes of $\Delta x, \Delta y = \xi$. This point onwards, the superscript "*" of the nondimensional parameters will be dropped for ease of representation of the results.

## III. RESULTS AND DISCUSSIONS

In this section, we focus on interfacial motion of immiscible fluids and its dependence on various chemical and electrical parameters of the binary system. Although a number of parameters of this coupled multiphysics problem can directly and/or indirectly affect the fluid transport behavior, here, we consider the following significant key parameters: (a) wettability contrast ($w_c$) which is the ratio of the static contact angles $\theta_s$ (wettability conditions on the patches) and is expressed by $w_c = \theta_B / \theta_A$; (b) width of the patches ($p_w$); (c) frequency ratio ($\omega$). We specify static contact angle on the different patches to impose surface affinity condition, the phase field modeling framework allows the interface to adjust dynamically in tune with the dynamic contact angle as the interface traverses along the channel. Some dimensionless parameters are kept constant throughout the analysis, such as: $Pe_\phi = 0.01$, $Pe_T = 0.07$, $Re = 0.01$, $Ca = 0.1$, $B = \eta_T / \eta_0 = -0.02 K^{-1}$, $\zeta = 50$, and $J = 1$. Moreover, we consider the property ratios as unity, such that the value of the properties for the two phase are equal. KCl electrolyte solution is taken as base fluid (fluid A) and dimensionless parameters are evaluated based on property values of fluid A. The widths of the different patches are equal and wettability conditions at same axial location are different for top and bottom walls to maintain the asymmetric surface patterning. In addition, the various dimensions shown in the Fig. 1 are $H = 1, L = 5, L_1 = 1$, $L_2 = 3, d_1 = 0.1, d_2 = 0.4, g = 0.1$ and $s = 0.2$.

Based on literature review on interfacial dynamics of immiscible fluids over wetted surfaces, it is inferred that the interfacial motion and associated contact line dynamics are well described by time sequences of interface profile, contact line velocity and filling time. Therefore, to delineate the thorough investigations on intricate physical characteristics of interfacial motion and its control over specific locations we have presented the results in the form of interface profile, contact line velocity and filling tine.

When immiscible binary fluids flow over patterned surface, the wetting characteristics of the patterned surface can impart stick-slip behavior of the contact line depending on the affinity conditions of the surface patterning. In order to observe the influence of the wettability contrast ($w_r$) on the flow control of the immiscible binary fluids we show the time evolutions of the interface profile and corresponding contact line velocity ($V_{cl}$) with axial position of interface in Fig. 3 for three different contrasts, $r = 1.25, 2$ and $3.5$, corresponding static contact angles of the patches are $\theta_A = 80°, 60°, 40°$ and $\theta_B = 100°, 120°, 140°$, respectively. The patch width $p_w = 0.5$ is kept constant for all the cases, while the other relevant other parameters are mentioned in the figure caption. The contact line velocity on bottom wall is denoted by $V_{cl,b}$ while $V_{cl,t}$ denotes the contact line velocity on the top wall. From the figure, it is clear that the interface contours are loosely packed over the A-type (green color) stripes whereas interfaces become denser over the B-type (red color) stripes. This can be attributed to the fact that A-type stripe has strong affinity for the fluid A (displacing fluid) whereas it dislikes the fluid B (displaced fluid). Therefore, as the interface traverses over the A-type regions, it experiences slip behavior. In contrast, B patches favorably attract the displaced fluid and hence, the motion of the binary fluids gets resisted and the interface shows sticky behavior on the B-type regions. The stick-slip behavior of the interface can also be explained differently. The surface tension force aids the electrical force to pull the interface over the A-type stripes for the predefined wettability condition $\theta_s < 90°$. On the other hand, two forces oppose with each other on the B-type stripes where $\theta_s > 90°$. From the interface plots, one can see that due to asymmetric patterning of the patches, the two ends of the interface touch patterned regions of two different kinds: one of which aids the movement whereas the other one opposes. Therefore, when one end moves at faster rate, another end moves gently and can even experience pinning near the junction of the stripes. Thus, the desirable interfacial motions with interface pinning are well controllable at two different locations (on top and bottom wall). Further, the flow characteristics and the delaying time depend significantly on the wettability contrast. Higher the wettability contrast, higher will be the difference of surface free energy between the adjacent patches. From the figure it is evident that as the wettability contrast increases, the areas of crowding of the interface profiles gradually become denser over the red patches. Therefore, flow can be decelerated significantly, and it can even be stopped with the alteration of the wettability contrast. Another interesting point of our analysis is that after a sufficient delaying time, flow can be accelerated to move to the next station where again it experiences deceleration and the flow can momentarily stop (see the interface profile of the Fig. 3(c)). The conditions apparent from such fluid motion seem like a stop-go valve to control the interfacial dynamics of the immiscible fluids.

The dynamical behavior of the interfacial motion of the binary fluids can be described in a more involved way by investigating contact line velocity plots. From the velocity plots, it can be observed that fluctuation of the contact line velocity is more prominent at higher wettability contrast. The fluctuation in contact line velocity can be attributed to the alternating changes in the direction of surface tension force and its interaction with ACET forces as the interface traverses along the channel. Large difference of free surface energy of the adjacent stripes, at higher wettability contrast, leads to relatively high peak velocities. The surface tension force may become equally dominating and when it opposes the driving force (i.e., ACET forces) the fluid-fluid interface stops at specific locations. Even sometimes, contact line/interface starts retreating and reverse movement takes place at relatively higher contrast in wettability, as can be seen from the inset to Fig. 3(b) and 3(c). Therefore, alternate stripes act as microvalves and the characteristics of the microvalve system depends on the properties of the stripes.

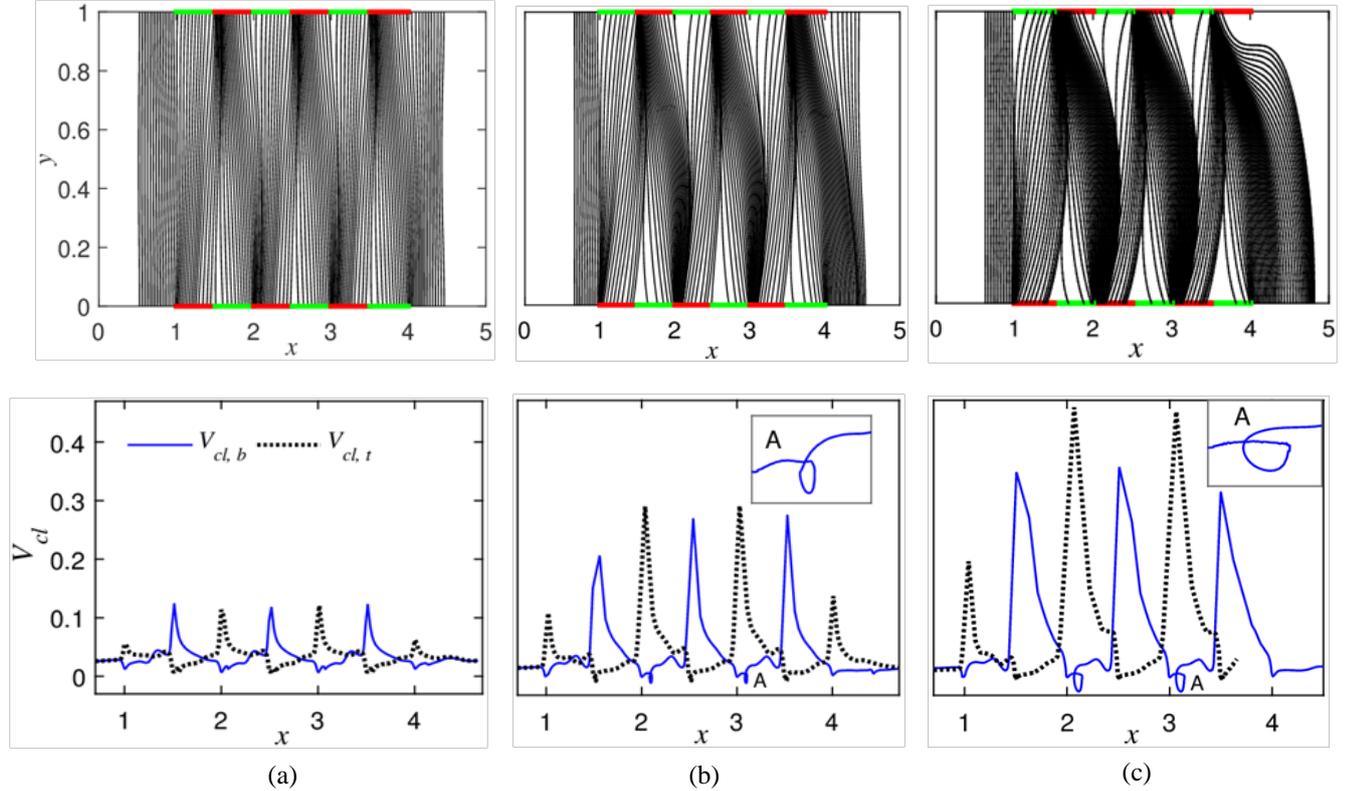

(a)  (b)  (c)

FIG. 3. Time evolutions of interface and associated contact line velocity plots with interface location for (a) $w_c = 1.25$, $\theta_A = 80°$, $\theta_B = 100°$ ; (b) $w_c = 2$, $\theta_A = 60°$, $\theta_B = 120°$ ; (c) $w_c = 3.5$, $\theta_A = 40°$, $\theta_B = 140°$. The other parameters are $p_w = 0.5$, $\zeta = 50$ and $\omega = 0.001$. The interface profiles follow stick and slip motion over red-type and green-type patches, respectively. Surface affinity conditions on red patches try to attract the displaced fluid thereby contact gets pinned over these patches. With increasing $w_c$, strength of the pinning-depinning action becomes stronger and delaying time of interface increases. Sudden change in free surface energy is attributed with sudden change in velocity. With increasing $w_c$, peaks of the velocities increase.

In the previous subsection, we have discussed the interplay of ACET forces and surface tension force in dictating the stick-slip behavior of the contact line which is eventually manifested as the accelerating-decelerating dynamics of the interface. Higher contrast in the wettabilities (dictated by $\theta_s$) of adjacent stripes leads to higher magnitudes of acceleration/deceleration of the interface. By suitably controlling wettability contrast, it is possible to momentarily stop the interface at some specific locations of the channel. However, the extent of delaying time that can be achieved for a specified wettability contrast, was not discussed. Next, we focus on the delaying time as modulated by $w_c$, at different locations in the channel. In Fig. 4 we show the filling time with axial distance of bottom and top walls for different wettability contrasts for a patch width of $p_w = 0.5$, other parameters used in the results are same as shown in the caption of the Fig. 3. From the figure, it can be seen that at specific locations $x = 1, 2, 3, 4$ for bottom wall and $x = 1.5, 2.5, 3.5$ for top wall, interface shows pinning and thus delaying time. However, delaying time is smaller at lower wettability contrast. Typically, for $w_c = 1.25$ it is almost negligible. Owing to lower difference in surface energy between the adjacent stripes the resistive force for stopping the motion at specific positions is less, and thus, the interface gradually progresses through the channel without delaying. On the other hand, for $w_c = 3.5$ a much longer delaying time is observed (see inset figure of Fig. 4(a)). Therefore, performance of the stop-go valve system is more effective at higher wettability contrast.

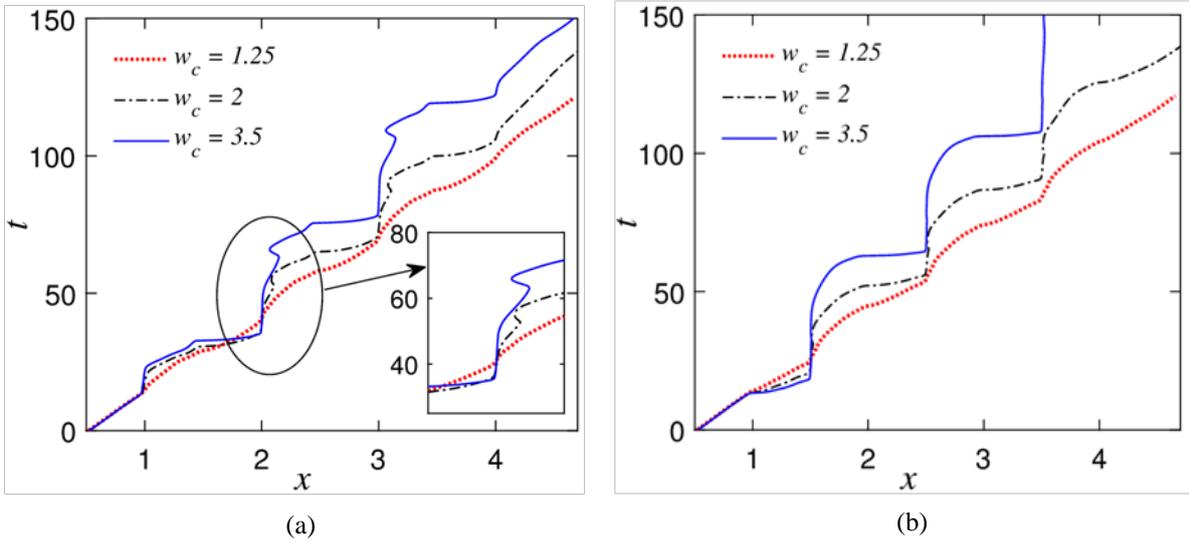

(a)    (b)

FIG. 4. Effect of wettability contrast. Time vs interface position over (a) bottom wall and (b) top wall. For both the walls, at lower wettability contrast ($w_c$) oppositely acting surface tension force loses its value, and, thus, the delaying times at locations $x = 1, 2, 3$ and 4 for bottom wall and $x = 1.5, 2.5$ and 3.5 at top wall drop noticeably. The inset figure of bottom wall shows the details of delaying time at location $x = 2$.

Another passive way to control the stop-go motion of the binary system is the alteration of width of the surface patches. In an effort to highlight the effect of the patch width the interface contours and corresponding velocities with channel length are shown in the Fig. 5 for three different patch widths of $p_w = 0.125, 0.25$ and $0.5$, other relevant parameters are mentioned in the figure caption. It is worth mentioning here that since the region of the observation section i.e., $L_2$, is kept fixed, the number of patches increase with decrease in patch width and vice versa. Therefore, the number of pinning stations, i.e., locations where the flow motion will stop, increases with decrease in $p_w$. It can be seen from the figure that for all the patch widths, the regions of crowding of the interface are observed over the red patches. Therefore, with decrease in patch width, the interface experiences frequent pinning states. However, a closer look on interface profiles reveals that the area over which the interfaces are closely packed is decreased with the increment of the number of patches. As a result, the delaying time over the halting locations is reduced. Although the wetting conditions on the patterned substrates are identical, due to lesser patch width, the retention time of the interface over the B patches is less. Hence the interface cannot stay for a long time. Quite notably, periodic alteration of the free surface energy for different patches has a significant effect on the interface profile. For larger patch width ($p_w = 0.5$), the sticking action of the B patches stops the interface for longer time while slippery patches (A patches) assist downstream motion of the interface. Therefore, larger interface stretching owing to asymmetric texturing of the surfaces is observed. In contrast, for smaller patch width ($p_w = 0.125$) stick-slip motion prevails for short time and thus the interface stretching is relatively low.

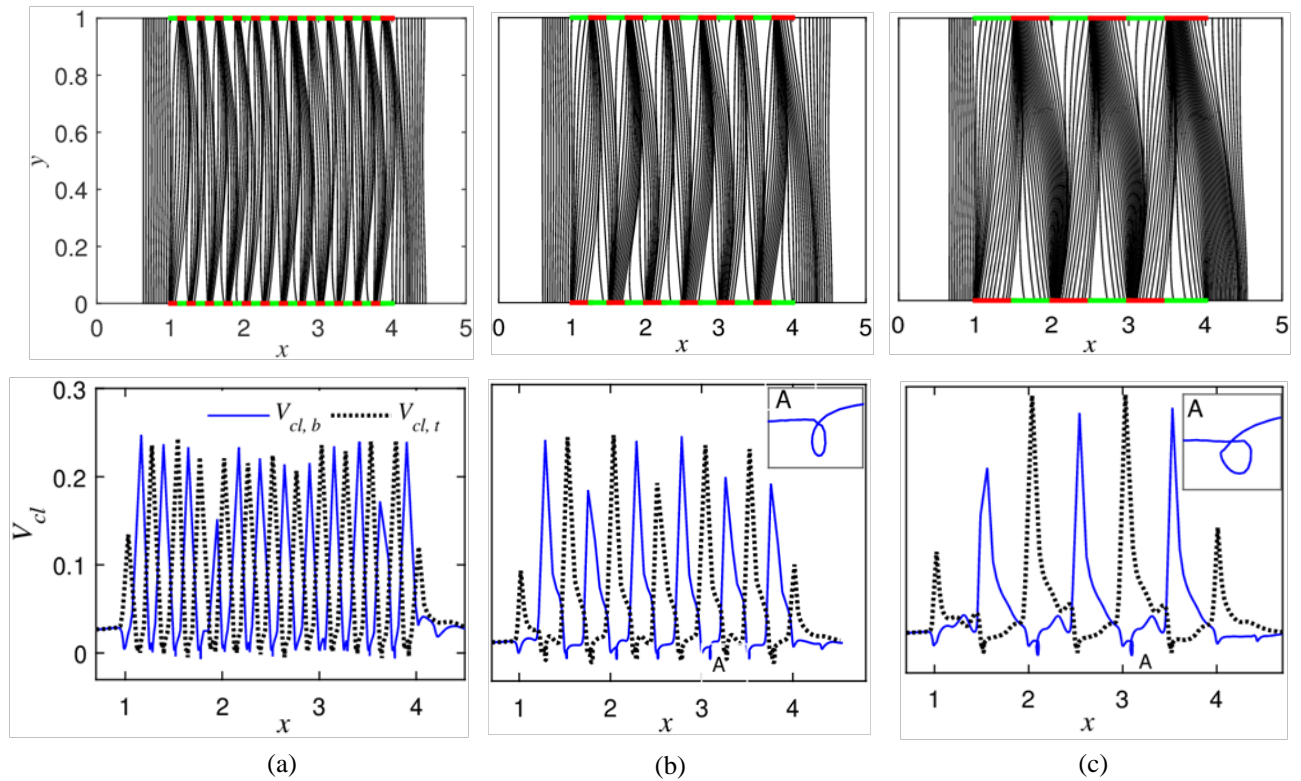

(a)  (b)  (c)

FIG. 5. Time sequences of interface profile and contact line velocity with interface location for (a) $p_w = 0.125$; (b) $p_w = 0.25$; (c) $p_w = 0.5$, the other parameters used in the results are $w_c = 2$, $\zeta = 50$ and $\omega = 0.001$. For a fixed patterned length, with increasing patch width number of patches reduces, but the interface can stop for a long time. From the inset figure of the velocity one can see the enlarge view of the interface retreating motion.

The alteration of the two states: halting and then faster movement from one station to next station of the valving system with changing patch width will be further explored by observing contact line velocity plots (Fig. 5). It is evident from the figure that according to the altered patch width (or number of patches) the contact line velocities are modulated over different wettability conditions. For example, for patch width of $p_w = 0.125$ the number of patch and number of velocity peaks are same (i.e., $24$). Sudden change in surface energy at the junctions of the two different patches results abrupt alteration of the net force across the contact line and thereby increases the contact line velocity. However, identical affinity conditions (manifested by static contact angle) between the adjacent wetted patches of three different patch widths result in almost equal (in magnitude) velocity peaks on the bottom and top walls. However, delaying times of the interface are not same over the patches for different widths. The velocity minima ($\approx 0$) of the contact line velocity show longer halting for the larger patch width (see the inset figures of Fig. 5(b) and 5(c) of velocity plots). The surface tension force which opposes the driving ACET forces is effective for a long time over the B-type stripes for largest patch (i.e., $p_w = 0.5$). However, as we decrease the patch width, although the number of regions of halting increasing the delaying time decreases owing to short time activation of the oppositely acting surface tension force.

We also represent the delaying time of the interface over the bottom and top walls for different patch widths in Fig. 6(a) and 6(b), respectively. The other parameters used in the plots are same as mentioned in the caption of Fig. 5. It is evident that halting locations become double on changing the patch width from $p_w = 0.5$ to $p_w = 0.25$ for both bottom and top walls. Decrease in patch widths for same observation distance ($L_2$) the number of patches increases. Therefore, number of red patches responsible for stopping the flow motion increases. As a results, the number of stations of halting increases. However, halting time significantly reduces for shorter patch widths. This is obvious, because area over which the sticking motion occurs become shorter on decrement of the patch width. Therefore, span of delay-time is also short. All the important massages are clearly visualized in the inset figures.

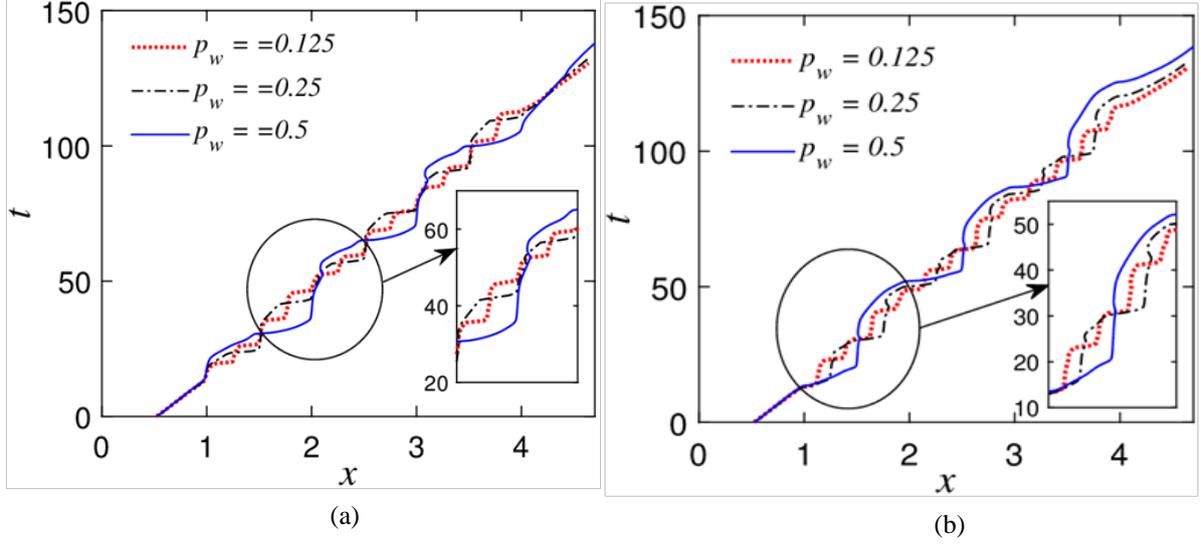

FIG. 6. Effect of patch width. Filling time vs axial distance over (a) bottom wall and (b) top wall. The inset figures show enlarge view of time at location $x = 2$ and $x = 1.5$ for bottom and top walls, respectively. For both the walls, it is clear that number of locations is increased with decreasing patch width. However, the stopping time at those locations appreciably decreases.

In the previous subsections, we have discussed how the surface property and patterning can be effectively used to passively control the interfacial motion of the immiscible fluids in a microvalving system. The surface tension force and its direction can be modulated by altering the wettability patterning of the surfaces. In our study, we employ electrothermal forces to actuate the two-phase motion. Applied frequency is one of the key parameter to alter performance of the active valving and the residence time that we have seen through our numerical investigations. From the general description of the electrothermal mechanism, it is seen that two components of the electrothermal force, namely Coulomb and dielectric forces (see Eq.(10)) are strong functions of the applied frequency [52]. Depending on the range of frequency, either or both of the force components may be significant. The directions of action of these forces are different, and hence, in the range of frequency when both forces are equally dominating, the net driving force becomes weaker. In the following section, we investigate the functional dependencies of these forces and the interplay of various components of ACET forces with the surface tension force which alter the functionality of the microvalving mechanism.

In an effort to investigate the flow control of a immiscible binary fluid system with altered AC frequency, we show the time evolutions of interface profile and related contact line velocity plots with position of the contact line in the Fig. 7 for $\omega = 0.001, 0.005$ and $0.01$; other relevant parameters are shown in the caption. Previously, from the investigations of electrothermal micropumping it was found that below the threshold applied frequency (order of $1\mathrm{MHz}$) pumping velocity was almost unaltered [46,57,72]. However, flow velocity gradually decreases as we increase frequency above this critical limit and becomes almost zero at cross over frequency ($\omega = \sigma/\varepsilon$). [60] In the present analysis, we have observed that below a threshold value of frequency ratio ($\omega = 0.001$) the contact line velocity was unchanged. In our numerical model, we consider reference frequency as $\omega_0 = \sigma_0/\varepsilon_0$. Typically, for $\sigma_0 \approx 0.1\,\text{S/m}$ and $\varepsilon_0 \approx 10^{-10}\,\text{C/V}$ m, the reference frequency becomes $\omega_0 \approx 10^9\,\text{Hz}$. Therefore, for $\omega = 0.001$, the applied frequency is of the order of 1MHz which is in good agreement with the previously reported observations. With increase in frequency above threshold value, the contact line velocity drops. We have mentioned earlier that various components of the ACET forces are frequency dependent. Below the threshold range of frequency, the Coulomb force is constant owing to the saturation of the free charges in the AC domain and it dominates over the dielectric force. However, at higher frequency ($\omega > 0.001$) electric charge density decreases. As a result, Coulomb force falls gradually and dielectric force interplays with the Coulomb force. Since these forces are mutually opposing, the net driving force decreases at higher frequency. Hence, the contact line velocity reduces. At lower net ACET force, forcing may not be sufficient to displace the interface beyond the patterned section. In the interface plots of Fig. 7(b) and Fig. 7(c) one can see that after a sufficient span of time ($t = 150$), interface reaches a distance of $x \approx 3$ and $x \approx 2$ on bottom wall, $x \approx 3.5$ and $x \approx 1.7$ on top wall for $\omega = 0.005$ and $\omega = 0.01$, respectively. It is worth noting that for these scenarios of interface halting, the contact line stops over the B-type stripes. As discussed previously, wetting characteristics of the surface patterning over the B-type stripe allows surface tension force to oppose the driving force. When the ACET forces become relatively low the opposing surface tension force dominates, the contact line gets pinned over the red patches (B-type stripes). Moreover, at high frequency ratio, for example $\omega = 0.01$, the interface may start retreating over the red patches which follows a backward movement as can be seen from the Fig. 7(c) (inset of the contact line velocity plot). Therefore, by controlling the frequency of the AC signal, it is possible to considerably increase the residence time of the fluid-fluid interface at specific locations. Also, we can accelerate the interface by changing the frequency. Thus, the valving response is much faster with applied frequency of the AC signal. Next, we explore the influence of the changing frequency ratio on the interface residence time.

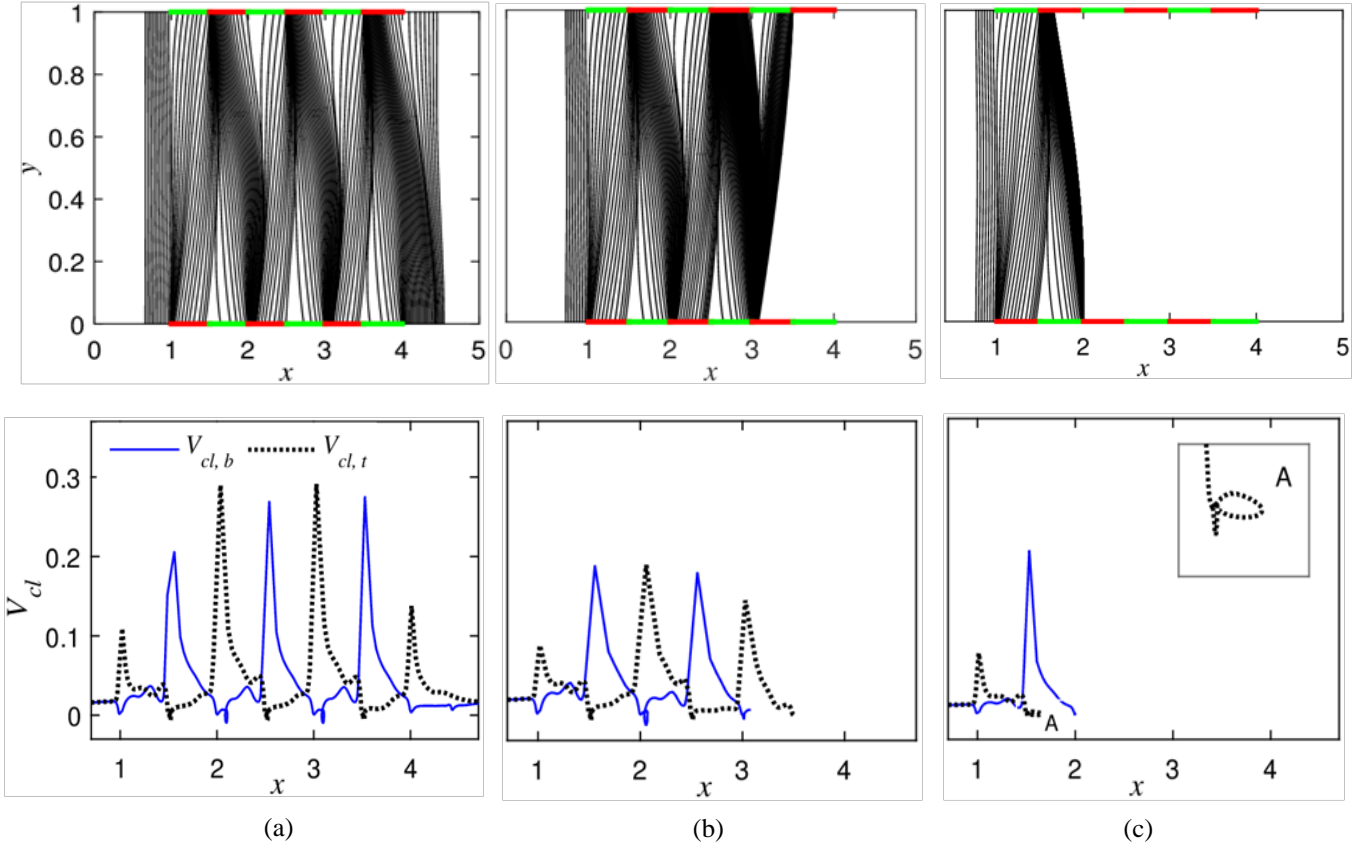

FIG. 7. Interface evolutions and relevant contact line velocity with contact line location for (a) $\omega = 0.001$ (b) $\omega = 0.005$ and (c) $\omega = 0.01$, the other parameters used in the numerical results are $w_c = 2$ $p_w = 0.5$ and $\zeta = 50$. The net ACET force is deeply affected with the changing applied frequency. The interaction between the Coulomb and dielectric force becomes significant at higher frequency ratio. Owing to opposite nature of action of these force driving force decreases at higher frequency ratio. Hence, interface is unable to cross the observation section although the elapse time is much enough ($t = 150$). Velocity plots also reveal that contact line velocity reduces with increasing frequency ratio. At high frequency interface start retreating motion over red stripes as can be seen from the inset figure.

Figure 8 shows the characteristics of delaying time of the interface with axial location for various frequency ratios, the other relevant parameters used in the simulations are mentioned in the caption of Fig. 7. It is evident that filling time and delaying time along the channel are identical for $\omega = 0.001$ and $0.0005$. As discussed previously, below the critical frequency $\omega = 0.001$, underlying physics of the interfacial motion is unaltered which is clearly depicted in the figure. Moreover, for $\omega = 0.005$, interface stops for a long interval at locations $x = 3$ and $x = 2.5, 3.5$ over the bottom wall and top wall, respectively. The reason is as stated above that at this frequency range, the dielectric force becomes significant and starts opposing the Coulomb force. As a result, the driving force becomes weaker and interface does not traverse the full observation section (i.e., patterned section). Further, with increase in frequency ratio ($\omega = 0.01$), the Coulomb and dielectric forces compete with each other. In this scenario, the interface gets pinned at $x = 2$ and $x = 1.5$ for bottom wall and top wall, respectively. B-type patch (red color) favorably attracts the B fluid. At lower net ACET forces, Coulomb, dielectric and surface tension forces are balanced and the movement of the interface was not found anymore. We can stop the fluids with our desired delaying time by changing the AC frequency of the active valving system.

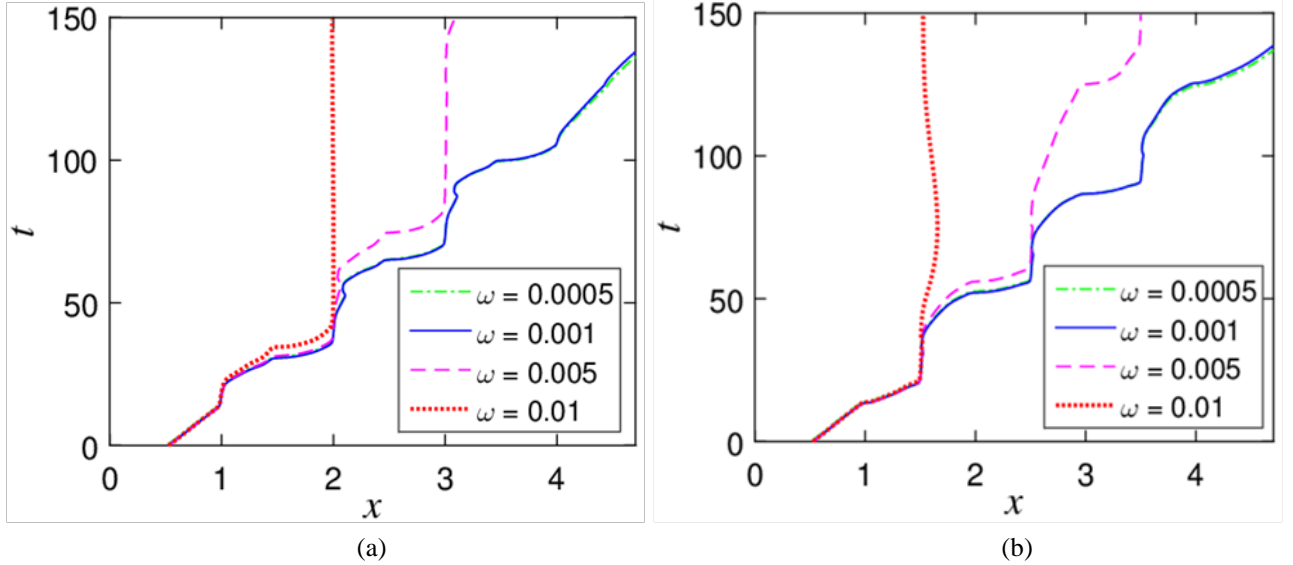

(a)             (b)

FIG. 8. Effect of frequency ratio . The variation of elapses time to traverse patterned length over (a) bottom wall and (b) top wall. The ACET mechanism is independent of frequency ratio for $\omega \leq 0.001$. So, variation of filling time is unaltered with . $\omega$ below the threshold value. However, at higher frequency gradual dominating behavior of dielectric force reduces the effective electrothermal force and the interface cannot move patterned length. For example, at $\omega = 0.01$ interface stop its motion at locations $x = 2$ and $x = 1.5$ for bottom wall and top wall, respectively.

## IV. CONCLUSIONS

We have investigated the performance of an active-passive microvalving system in a capillary whose surfaces are asymmetrically paved with two different wetted patches under AC electrokinetic phenomena. We focus on various characteristics of the channel surfaces and the relevant electrokinetic parameter which directly alter the effectiveness of the valving system. Accordingly, influences of the surface affinity conditions (manifested by static contact angle), width of the stripes, applied frequency are studied systematically. Interaction between the various components of the ACET forces and surface tension force can slow down the fluid speed, so that motion can even be stopped at desirable locations. The alternate wetted patches have strong affinities to attract either of the fluids. With increase in wettability contrast, difference of free surface energy between two adjacent patches increases and strong sticking action take places of the interface over B-like patches. Therefore, the delaying time significantly increases over B patches. Since, the displaced fluid gets pinned over the red stripes, decreasing in patch width reduces the residence time over the B patches, resulting drops in delaying time. The Coulomb force and dielectric force interplay with each other on alteration of the frequency. Hence, by controlling AC frequency, interfacial motion can be stopped, or accelerated at any time and location, at will. In addition, by switching on/off, the electrical signal, fluid motion can be stopped at specified locations in the channel. [92]

As a final note, we summarize that flow behavior of the two-phase binary system are strong functions of the microscale properties of the patterned surfaces and AC electrokinetics. By tuning the surface properties and the electrical parameters, we can develop a stop-go valving system. We believe that the fundamental insights drawn from the present analysis may help in developing and designing efficient microvalve to control the fluid flow which possesses broad applications in the field of bio-chemical and biomedical processing over miniaturized scales.

**References**

[1]   P. K. Mondal, U. Ghosh, A. Bandopadhyay, D. DasGupta, and S. Chakraborty, Soft Matter **10**, 8512 (2014).

[2]   B. Zhao, J. S. Moore, and D. J. Beebe, Science (80-. ). **291**, 1023 (2001).

[3]   H. Ding and P. D. M. Spelt, Phys. Rev. E **75**, 46708 (2007).

[4]   Y. D. Shikhmurzaev, Int. J. Multiph. Flow **19**, 589 (1993).

[5]   D. N. Sibley, N. Savva, and S. Kalliadasis, Phys. Fluids **24**, 82105 (2012).

[6]   Y. D. Shikhmurzaev, J. Fluid Mech. **334**, 211 (1997).

[7]   R. B. Schoch, J. Han, and P. Renaud, Rev. Mod. Phys. **80**, 839 (2008).

[8]   C. M. Lieber, MRS Bull. **28**, 486 (2003).

[9]   H. Craighead, Nature **442**, 387 (2006).

[10]  F. Mugele, Soft Matter **5**, 3377 (2009).

[11]  R. Dey, K. Raj M., N. Bhandaru, R. Mukherjee, and S. Chakraborty, Soft Matter **10**, 3451 (2014).

[12]  P. K. Mondal, D. DasGupta, and S. Chakraborty, Soft Matter **11**, 6692 (2015).

[13]  J. Yao, M. Yang, and Y. Duan, Chem. Rev. **114**, 6130 (2014).

[14]  C. L. Moraila-Martínez, M. A. Cabrerizo-Vílchez, and M. A. Rodríguez-Valverde, Soft Matter **9**, 1664 (2013).

[15]  P. Chen, M. T. Chung, W. McHugh, R. Nidetz, Y. Li, J. Fu, T. T. Cornell, T. P. Shanley, and K. Kurabayashi, ACS Nano **9**, 4173 (2015).

[16]  F.-C. Wang and H.-A. Wu, Soft Matter **9**, 5703 (2013).

[17]  A. W. Smith, Adv. Drug Deliv. Rev. **57**, 1539 (2005).

[18]  Y. Song, J. Hormes, and C. S. Kumar, Small **4**, 698 (2008).

[19]  A. Askounis, D. Orejon, V. Koutsos, K. Sefiane, and M. E. R. Shanahan, Soft Matter **7**, 4152 (2011).

[20]  P. K. Mondal, U. Ghosh, A. Bandopadhyay, D. Dasgupta, and S. Chakraborty, Phys. Rev. E **88**, 23022 (2013).

[21]  X.-P. Wang, T. Qian, and P. Sheng, J. Fluid Mech. **605**, 59 (2008).

[22]  P. K. Mondal, D. Dasgupta, A. Bandopadhyay, and S. Chakraborty, J. Appl. Phys. **116**, 84302 (2014).

[23]  O. Kuksenok, J. M. Yeomans, and A. C. Balazs, Phys. Rev. E **65**, 31502 (2002).

[24]  O. Kuksenok, D. Jasnow, and A. C. Balazs, Phys Rev E **68**, 51505 (2003).

[25]  O. Kuksenok, D. Jasnow, J. Yeomans, and A. C. Balazs, Phys. Rev. Lett. **91**, 108303 (2003).

[26]  N. M. Oliveira, A. I. Neto, and W. Song, Appl. Phys. Express **3**, 85205 (2010).

[27]  Z. Bai, Q. He, S. Huang, X. Hu, and H. Chen, Anal. Chim. Acta **767**, 97 (2013).

[28]  S. Karakare, A. Kar, A. Kumar, and S. Chakraborty, Phys. Rev. E **81**, 16324 (2010).

[29]  J. B. Pendry, Phys. Rev. Lett. **85**, 3966 (2000).

[30]  H. Logtenberg, M. J. Lopez-Martinez, B. L. Feringa, W. R. Browne, and E. Verpoorte, Lab Chip **11**, 2030 (2011).

[31]  P. Seppecher, Int. J. Eng. Sci. **34**, 977 (1996).

[32]  L. M. Pismen and Y. Pomeau, Phys. Rev. E **62**, 2480 (2000).

[33]  K. Tsougeni, D. Papageorgiou, A. Tserepi, and E. Gogolides, Lab Chip **10**, 462 (2010).

[34]  W.-G. Bae, S. M. Kim, S.-J. Choi, S. G. Oh, H. Yoon, K. Char, and K. Y. Suh, Adv. Mater. **26**, 2665 (2014).

[35]  K.-H. Chu, R. Xiao, and E. N. Wang, Nat. Mater. **9**, 413 (2010).

[36]  G. H. . Sanders and A. Manz, TrAC Trends Anal. Chem. **19**, 364 (2000).

[37]  D. Juncker, H. Schmid, U. Drechsler, H. Wolf, M. Wolf, B. Michel, N. de Rooij, and E. Delamarche, Anal. Chem. **74**, 6139 (2002).

[38]  A. Salari, M. Navi, and C. Dalton, Biomicrofluidics **9**, 14113 (2015).

[39]  J. J. Feng, S. Krishnamoorthy, and S. Sundaram, Biomicrofluidics **1**, 24102 (2007).

[40]  M. A. Unger, Science (80-. ). **288**, 113 (2000).

[41]  D. J. Harrison, K. Fluri, K. Seiler, Z. Fan, C. S. Effenhauser, and A. Manz, Science

(80-. ). **261**, 895 (1993).

[42] M. W. J. Prins, Science (80-. ). **291**, 277 (2001).

[43] K. M. Grant, J. W. Hemmert, and H. S. White, J. Am. Chem. Soc. **124**, 462 (2002).

[44] K. Yang and J. Wu, Biomicrofluidics **2**, 24101 (2008).

[45] J. Wu, M. Lian, and K. Yang, Appl. Phys. Lett. **90**, 234103 (2007).

[46] R. Zhang, C. Dalton, and G. A. Jullien, Microfluid. Nanofluidics **10**, 521 (2011).

[47] Q. Lang, Y. Ren, D. Hobson, Y. Tao, L. Hou, Y. Jia, Q. Hu, J. Liu, X. Zhao, and H. Jiang, Biomicrofluidics **10**, 64102 (2016).

[48] M.-H. Chang and G. M. Homsy, Phys. Fluids **17**, 74107 (2005).

[49] N. G. Green, A. Ramos, A. González, A. Castellanos, and H. Morgan, J. Phys. D. Appl. Phys. **33**, L13 (1999).

[50] D. Erickson, D. Sinton, and D. Li, Lab Chip **3**, 141 (2003).

[51] J. P. Urbanski, T. Thorsen, J. A. Levitan, and M. Z. Bazant, Appl. Phys. Lett. **89**, 143508 (2006).

[52] A. Ramos, H. Morgan, N. G. Green, and A. Castellanos, J. Phys. D. Appl. Phys. **31**, 2338 (1998).

[53] G. Kunti, A. Bhattacharya, and S. Chakraborty, J. Nonnewton. Fluid Mech. **247**, 123 (2017).

[54] M. Lian and J. Wu, Appl. Phys. Lett. **94**, 64101 (2009).

[55] E. Du and S. Manoochehri, Appl. Phys. Lett. **96**, 34102 (2010).

[56] N. Siva, K. Gunda, S. Bhattacharjee, and S. K. Mitra, Biomicrofluidics **6**, 34118 (2013).

[57] F. J. Hong, J. Cao, and P. Cheng, Int. Commun. Heat Mass Transf. **38**, 275 (2011).

[58] V. Studer, A. Pepin, Y. Chen, and A. Ajdari, Analyst **129**, 944 (2004).

[59] Hansen T.S., Simulation and Testing of AC Electroosmotic Micropumps, Master thesis, Technical University of Denmark, 2004.

[60] E. Du and S. Manoochehri, J. Appl. Phys. **104**, 64902 (2008).

[61] Y.-L. Chen and H.-R. Jiang, Biomicrofluidics **11**, 34102 (2017).

[62] G. M. Whitesides, Nature **442**, 368 (2006).

[63] M. Medina-Sánchez, S. Miserere, S. Marín, G. Aragay, and A. Merkoçi, Lab Chip **12**, 2000 (2012).

[64] S. Nagrath, L. V. Sequist, S. Maheswaran, D. W. Bell, D. Irimia, L. Ulkus, M. R.

Smith, E. L. Kwak, S. Digumarthy, A. Muzikansky, P. Ryan, U. J. Balis, R. G. Tompkins, D. A. Haber, and M. Toner, Nature **450**, 1235 (2007).

[65] J. M. Moreno and J. M. Quero, J. Micromechanics Microengineering **20**, 15005 (2010).

[66] H. Hartshorne, C. J. Backhouse, and W. E. Lee, Sensors Actuators B Chem. **99**, 592 (2004).

[67] W. H. Grover, R. H. C. Ivester, E. C. Jensen, and R. A. Mathies, Lab Chip **6**, 623 (2006).

[68] L. Gui and J. Liu, J. Micromechanics Microengineering **14**, 242 (2004).

[69] R. Pal, M. Yang, B. N. Johnson, D. T. Burke, and M. A. Burns, Anal. Chem. **76**, 3740 (2004).

[70] N. S. Satarkar, W. Zhang, R. E. Eitel, and J. Z. Hilt, Lab Chip **9**, 1773 (2009).

[71] Q. Lang, Y. Wu, Y. Ren, Y. Tao, L. Lei, and H. Jiang, ACS Appl. Mater. Interfaces **7**, 26792 (2015).

[72] F. J. Hong, F. Bai, and P. Cheng, Microfluid. Nanofluidics **13**, 411 (2012).

[73] D. Jacqmin, J. Fluid Mech. **402**, 57 (2000).

[74] V. E. Badalassi, H. D. Ceniceros, and S. Banerjee, J. Comput. Phys. **190**, 371 (2003).

[75] Y. Q. Zu and S. He, Phys. Rev. E **87**, 43301 (2013).

[76] X. Luo, X.-P. Wang, T. Qian, and P. Sheng, Solid State Commun. **139**, 623 (2006).

[77] Y. Y. Yan and Y. Q. Zu, J. Comput. Phys. **227**, 763 (2007).

[78] S. Mandal, U. Ghosh, A. Bandopadhyay, and S. Chakraborty, J. Fluid Mech. **776**, 390 (2015).

[79] D. Jacqmin, J. Comput. Phys. **155**, 96 (1999).

[80] G. Kunti, A. Bhattacharya, and S. Chakraborty, Phys. Fluids **29**, 82009 (2017).

[81] T. Qian, X.-P. Wang, and P. Sheng, Phys. Rev. E **68**, 16306 (2003).

[82] L. D.R., *CRC Handbook of Chemistry and Physics*, 84th ed. (CRC Press, London, 2003).

[83] H. Morgan and N. G. Green, *AC Electrokinetics: Colloids and Nanoparticles* (Research Studies Press, Philadelphia, 2003).

[84] A. Ramos, *Electrokinetics and Electrohydrodynamics in Microsystems* (Springer, New York, 2011).

[85] Castellanos A, *Electrohydrodynamics* (Springer, New York, 1998).

[86] H. Ding, P. D. M. Spelt, and C. Shu, J. Comput. Phys. **226**, 2078 (2007).

[87] Q. Li, K. H. Luo, Y. J. Gao, and Y. L. He, Phys. Rev. E **85**, 26704 (2012).

[88] N. G. Green, A. Ramos, A. Gonzalez, A. Castellanos, and H. Morgan, J. Electrostat. **53**, 71 (2001).

[89] H. Liu, A. J. Valocchi, Y. Zhang, and Q. Kang, J. Comput. Phys. **256**, 334 (2014).

[90] R. G. Cox, J Fluid Mech **168**, 169 (1986).

[91] O. V. Voinov, Fluid Dyn. **11**, 714 (1976).

[92] A. Castellanos, A. Ramos, A. González, N. G. Green, and H. Morgan, J. Phys. D. Appl. Phys. **36**, 2584 (2003).